\documentclass[aip,reprint,graphicx]{revtex4-1}
\usepackage{amssymb,amsmath,bm,graphicx}
\usepackage{color}
\usepackage[english]{babel}
\draft 

\begin{document}


\title{Influence of inherent structure shear stress of supercooled liquids on their shear moduli} 



\author{Ingo Fuereder}
\author{Patrick Ilg}
\altaffiliation[Also at ]{School of Mathematical and Physical Sciences, University of Reading, Reading RG6 6AX}

\affiliation{ETH Zurich, Department of Materials, Vladimir-Prelog-Weg 1-5/10, CH-8093 Zurich}


\date{\today}

\begin{abstract}
\noindent
Configurations of supercooled liquids residing in their local potential minimum (i.e.\ in their inherent structure, IS) were found to support a non-zero shear stress. This IS stress was attributed to the constraint to the energy minimization imposed by boundary conditions, which keep size and shape of the simulation cell fixed. In this paper we further investigate the influence of these boundary conditions on the IS stress. We investigate its importance for the computation of the low frequency shear modulus of a glass obtaining a consistent picture for the low- and high frequency shear moduli over the full temperature range. Hence, we find that the IS stress corresponds to a non-thermal contribution to the fluctuation term in the Born-Green expression. This leads to an unphysical divergence of the moduli in the low temperature limit if no proper correction for this term is applied. Furthermore, we clarify the IS stress dependence on the system size and put its origin on a more formal basis.
\end{abstract}


\maketitle 

\section{Introduction}
\noindent
Upon decreasing temperature supercooled liquids display a dramatic increase in their shear viscosity. This effect can be directly related to an increase of the relaxation time of shear stress fluctuations. In his seminal work Goldstein \cite{Goldstein} pointed out the importance of the potential energy landscape over the $N$ particle configuration for the slow relaxation dynamics of supercooled liquids. At a sufficiently low temperature the system is assumed to be close to a local minimum of this landscape (called an inherent structure).\\

This picture naturally implies the presence of two time scales in supercooled liquids: a fast relaxation process associated with vibrational motion of the system around an inherent structure and a slow relaxation process corresponding to thermally activated hopping to another minimum in the potential energy landscape accompanied by a rearrangement of a relatively small number of particles in the system.
In a non-equilibrium situation applied strain might induce the disappearance of a local minimum facilitating such a rearrangement and leading to stress relaxation.
While this fact has motivated investigations on the influence of strain on the local potential minima \cite{Lacks,AQS,majid,karmakar}, several studies have implicitly made use of residual stresses in inherent structures present even at equilibrium investigating the magnitude \cite{magn} and the relaxation dynamics \cite{ViscNetwork,stressrelax} of the inherent stresses as the system samples different minima. As pointed out by Abraham and Harrowell, the notion of a shear stress of an inherent structure is a priori far from obvious: \cite{Harro} if a system which can be brought arbitrarily close to a local energy minimum (say by an appropriate, numerical energy minimization of a simulation), one would naively expect that all global shear stresses in the system are zeroed by this minimization procedure. However, it has been argued that boundary conditions impose a constraint on the energy minimization i.e.\ the shape and the size of the containment of a supercooled liquid forbids to zero all stresses. In a computer simulation this constraint amounts to a particular choice for the simulation box.
It is not clear in which sense this IS shear stress can be considered as a predecessor of the stresses supported by a deformed supercooled liquid/glass or to which extent it determines the stress relaxation process in a non-equilibrium situation at all. Therefore, a better understanding of the IS stress is highly desirable. A first discussion of it has also been given in [\onlinecite{Harro}]. Among other things, the authors found that the magnitude of the IS stress is surprisingly, essentially independent of temperature, scales with a certain power of the system density and with the inverse system size. \\
The aim of this note is to lift the inherent structure stress on a more formal footing which will help us to further clarify its origin and to discuss in what sense it has an influence on computations of viscoelastic properties. To this end, we will present a consistent picture for the proper calculation of low and high frequency shear moduli of glass-forming liquids. This work is organized as follows: In section II we begin our discussion by applying the Irving-Kirkwood formula for IS configurations and tracing back the remaining stresses to the choice of boundary conditions in a formal sense. In section III, we proceed by identifying external mechanisms which bias the computation of shear moduli upon decreasing temperature in a supercooled liquid and provide calculations of these moduli for glass forming liquids. In section IV we summarize our results and conclude with a discussion of the physical meaning of the IS stresses.


\section{The Irving-Kirkwood formula for Inherent Structure configurations}
\noindent
According to Irving and Kirkwood, the instantaneous stress tensor in a configuration with particle mass $m_{i}$, positions $\bm{r}_{i}$ restricted to a volume $V$ is given by \cite{irving}
\begin{equation}
\label{IK1}
     \sigma_{\alpha \beta}=\frac{1}{V} \sum_{i=1}^{N} m_{i} v_{i,\alpha} v_{i,\beta} - \frac{1}{2V} \sum_{i=1}^{N} \sum_{j=1}^{N} r_{i j,\alpha} F_{i j,\beta}   \ ,
\end{equation}
where $\bm{r}_{ij}=\bm{r}_{i}-\bm{r}_{j}$, $\bm{v}_{i}=\dot{\bm{r}}_{i}$ and $\bm{F}_{ij}$ is the pair forces exerted on particle $i$ by particle $j$. The Greek indices refer to the cartesian component of the corresponding vector/tensor. In an IS configuration (i.e.\ in a mechanically stable packing) the particles are at rest and the inherent structure stress $\sigma^{IS}$ is solely determined by the second part of ($\ref{IK1}$).
We assume periodic boundary conditions in a cubic box with side length $L$ i.e.\ we redefine $r_{ij,\alpha}=r_{i,\alpha}-r_{j,\alpha}+n_{ij,\alpha} L$. Here, $\bm{n}_{ij}$ denotes the vector which minimizes the distance between particles $i$ and $j$ where its components are $\pm 1$ or $0$. Inserting the periodic boundary conditions in the configurational part of ($\ref{IK1}$) leads to
\begin{multline}
\label{IK2}
     \sigma_{\alpha \beta}^{IS}= - \frac{1}{2V} \left( \sum_{i=1}^{N} r^{IS}_{i,\alpha} \sum_{j=1}^{N} F_{i j,\beta} \right. \\ \left. - \sum_{j=1}^{N} r^{IS}_{j,\alpha} \sum_{i=1}^{N} F_{i j,\beta}  + L \sum_{i=1}^{N} \sum_{j=1}^{N} n_{ij,\alpha} F_{i j,\beta} \right)   \ .
\end{multline}
By using Newton's third law $\bm{F}_{ij}=-\bm{F}_{ji}$ and renaming of indices, this can be rewritten as follows.
\begin{equation}
\label{IK3}
     \sigma_{\alpha \beta}^{IS}= - \frac{1}{V} \sum_{i=1}^{N} r^{IS}_{i,\alpha} F_{i,\beta} - \frac{L}{2V} \sum_{i=1}^{N} \sum_{j=1}^{N} n_{ij,\alpha} F_{i j,\beta}   \ ,
\end{equation}
where $\bm{F}_{i}=\sum_{j=1}^{N} \bm{F}_{i j}$ is the net force acting on particle $i$. This quantity vanishes by definition for an IS. In a simulation it is arbitrarily small in the sense that the magnitude of the net force on a particle is bounded by the smallest force tolerance at which the used minimization algorithm (e.g.\ a conjugate gradient solver) still converges. Therefore, the IS stress is approximately given by
\begin{equation}
\label{IK4}
     \sigma_{\alpha \beta}^{IS} \approx  - \frac{L}{2V} \sum_{i=1}^{N} \sum_{j=1}^{N} n_{ij,\alpha} F_{i j,\beta}   \ .
\end{equation}
 \begin{figure}
 \includegraphics[scale=0.6]{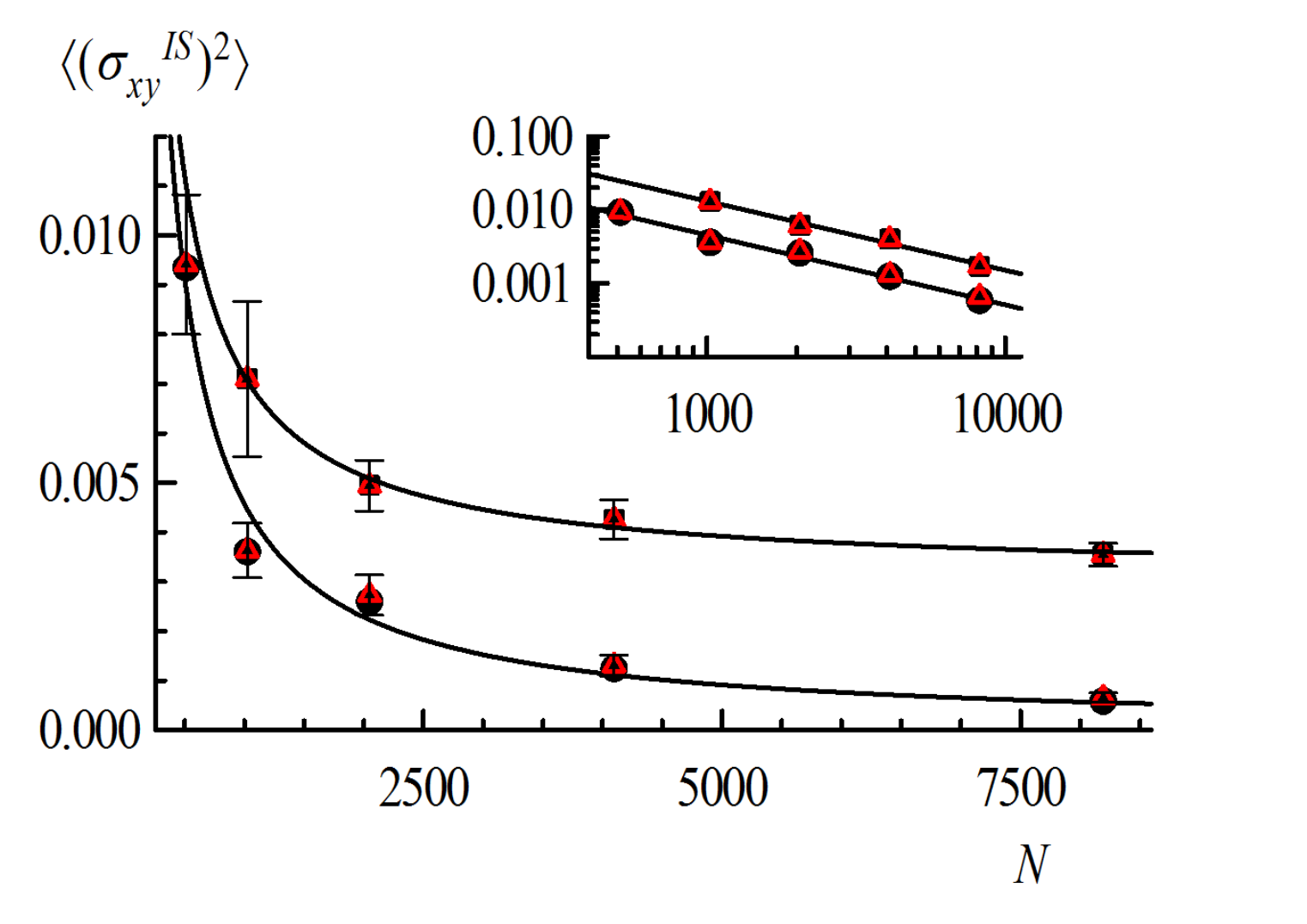}%
 \caption{\label{fig1} Variance of inherent structure stress in 2D (top, soft sphere system) and 3D (bottom, binary LJ system) both at temperature T=0.5 as a function of system size. Black circles indicate the full Irving-Kirkwood expression and red triangles the approximate boundary formula ($\ref{IK4}$). The insets show the same data on a $\log$-$\log$ scale with a linear fit which has a slope of $-1$. }
 \end{figure}
As we will discuss subsequently, expression ($\ref{IK4}$) has a straightforward physical interpretation. Note, that the $\alpha$-component of the vector $\bm{n}_{ij}$ is only nonzero if a particle close to the $\alpha=L/2$ boundary of the simulation box interacts with the periodic image of a particle residing at the opposite ($\alpha=-L/2$) boundary or vice versa. For instance, if we choose $\alpha=x$, $\beta=y$ in a cartesian coordinate system, expression ($\ref{IK4}$) is nothing else than the total force component in $y$-direction exerted by the right boundary layer of the simulation box on its left counterpart. This shows analytically in which sense the choice of boundary conditions (i.e.\ the shape of the simulation box) determines the IS stress. We test approximation ($\ref{IK4}$) numerically by comparing the mean squared IS stress calculated from the full Irving-Kirkwood expression to the one calculated from approximation ($\ref{IK4}$) by performing molecular dynamics simulation of glass forming systems for different system sizes, temperatures and dimensions (see appendix A for details on the simulations). The total inherent structure stress is very well approximated by equation ($\ref{IK4}$) (see Fig.1).
As it can be read off from Fig. 1, the fluctuations of $\sigma_{\alpha \beta}^{IS}$ show a $1/V$ system size dependence, which has already been described described in [\onlinecite{Harro}]. A heuristical explanation is given in appendix C based on equation ($\ref{IK4}$).
The investigation of the IS in amorphous materials dates back to ideas of Stillinger and Weber \cite{stillinger} and has proven to have various applications including the investigation of rate processes in low-temperature amorphous substances, the formulation of equations of state for supercooled liquids, macroscopic transport properties etc. (see e.g.\ [\onlinecite{sciortinoI}],[\onlinecite{sciortinoII}] and [\onlinecite{heuerrev}]). While equation ($\ref{IK4}$) seems to identify the IS stress to be a mere (negligible) boundary effect, we will discuss that it is felt throughout the system and has a major influence on macroscopic quantities. Our discussion will focus on the influence of IS stresses on the numerical computation of elastic constants, more specifically on the shear moduli of a glass forming material.

\section{Elastic constants}
\noindent
The shear modulus, $G$, describes the response of a material to shear stress and is defined as the ratio between shear stress to shear strain. If a material is subjected to oscillatory shear deformation, the shear modulus is a function of the excitation frequency $\omega$, i.e.\ $G=G(\omega)$. Its high- and low frequency limits are characteristic for a material's mechanical behavior: the infinite frequency shear modulus, $G_{\infty}$, describes the response to an instantaneous, affine deformation. It is not directly measurable experimentally, since a deformation at truly infinite frequency cannot be applied in practice. $G_{\infty}$ should not be confused with the experimentally reported high frequency modulus which always refers to the shear modulus at the highest obtainable frequencies. \cite{dyre} The temperature dependence of $G_{\infty}$ is relatively weak and its value mainly depends on the microscopic details (atomistic potential) of the system. The zero frequency limit, $G_{0}$, describes the ability of relaxing stresses on a long time scale. Since a liquid in equilibrium does not support any stresses, $G_{0}$ is zero for a liquid in equilibrium but finite for a solid. Therefore, the zero frequency modulus can be regarded as an indicator for solidity. It depends strongly on the thermodynamic state and is therefore sensitive to temperature changes as a material approaches its melting point. The situation is more complicated for glassforming materials as they do not show a sharp solidification transition. In the following, we will summarize and extend previous results for $G_{\infty}$ and $G_{0}$ of a supercooled liquid and discuss their behavior over the full temperature range. Furthermore, we will investigate the contribution stemming from the IS and discuss in what sense it contributes to properties of the low temperature glass. Throughout this section the simulation results are reported for the two-dimensional soft sphere system with $N=512$ particles (see appendix).

\subsection{General remarks}
\noindent
The infinite frequency shear modulus is analytically given by the so-called Born-Green expression \cite{born} which is well defined and yields non-vanishing results in both the solid and the fluid phase.
This expression is given by

\begin{multline}
\label{BG1}
     G_{\infty}=\rho k_{B} T+ \\ \left \langle \frac{1}{2V}\sum_{i,j}r_{ij,x}^{2} r_{ij,y}^{2} \left( \frac{\phi''(r_{ij})}{r_{ij}^{2}}-\frac{\phi'(r_{ij})}{r_{ij}^{3}}\right) \right\rangle -P    \ ,
\end{multline}
for an \textit{isotropic} system with the hydrostatic pressure $P$ and a pair potential $\phi(r)$. Further utilizing isotropy and performing an orientational average this can be simplified to  \cite{zwanzig, hess}
\begin{equation}
\label{BG2}
     G_{\infty}=\rho k_{B} T + \frac{1}{8V}\left\langle\frac{1}{2}\sum_{i,j} \frac{1}{r_{ij}} \frac{\partial}{\partial r_{ij}} \left[r_{ij}^{3}\phi'(r_{ij}) \right]  \right\rangle   \ ,
\end{equation}
for a two dimensional system. For a soft sphere system with a purely repulsive pair potential of the type $r^{-n}$ equation ($\ref{BG2}$) leads to \cite{ilyin}
\begin{equation}
\label{BG3}
     G_{\infty}=\rho k_{B} T + \frac{n-2}{4}(P-\rho k_{B} T)   \ .
\end{equation}
The low frequency shear modulus is given by equation ($\ref{BG1}$) corrected by the so-called fluctuation term \cite{Squire,MM,hess}, i.e.
\begin{equation}
\label{BG4}
     G_{0}=G_{\infty} - \frac{V}{k_{B} T} \left( \left\langle \sigma_{xy}^2 \right\rangle  - \left\langle \sigma_{xy} \right\rangle^{2}  \right) \ .
\end{equation}
In the following we will provide an extensive discussion of these quantities for a glass forming liquid in different temperature regimes.

\subsection{High temperature liquid}
\noindent
For temperatures well above the melting point of the system, the low frequency shear modulus vanishes, i.e.\ the system does not sustain non-zero stresses on a long time scale. In this situation the fluctuation term cancels the Green-Born expression and the high frequency shear modulus can be calculated by the shear stress fluctuations:
\begin{equation}
\label{BG5}
     G^{L}_{\infty}= \frac{V}{k_{B} T} \left( \left\langle \sigma_{xy}^2 \right\rangle  - \left\langle \sigma_{xy} \right\rangle^{2}  \right) \ .
\end{equation}

Note that the ensemble average of the shear stress $\left\langle \sigma_{xy} \right\rangle$ in equilibrium always vanishes.

\subsection{Supercooled phase}
\noindent
Upon further cooling the system below its melting point, the supercooled regime is entered. The fluid is not in its true thermodynamic equilibrium being the crystalline phase but is said to be in a metastable equilibrium in the sense that time translational invariance holds and two time correlation functions (such as the stress-stress autocorrelation function, $C(t)=\left \langle \sigma_{xy}(t)\sigma_{xy}(0)\right\rangle$) decay to zero within experimentally available time windows (see fig.\ 2). As the particle motion becomes increasingly sluggish, relaxation becomes slower and happens on two time scales: vibrational degrees of freedom lead to a fast redistribution of stresses (i.e.\ an initial decay of $C(t)$ to the lowest possible value possible for a particular configuration in place). This process is followed by a slow relaxation associated with particle rearrangements on a mesoscopic scale. As the system is still (quasi-)ergodic in the sense that it finds a way to redistribute stresses such that $C(t)$ fully decays to zero, the formulas for the elastic moduli (equations ($\ref{BG3}$) and ($\ref{BG4}$)) are still valid. However, since the relaxation of stress correlations takes an increasingly long time scale, the liquid assumes a viscoelastic behavior and the low frequency shear modulus departs from its zero value. According to equation ($\ref{BG4}$), this is associated with a decrease in the shear stress fluctuations. It also means that equation ($\ref{BG5}$) is not appropriate to compute the high frequency modulus anymore, but equation ($\ref{BG1}$) has to be used. The fact that the common expression ($\ref{BG5}$) looses its validity in the solid phase has been pointed out by several authors. \cite{Harro, WiliamsI, ilyin}
\subsection{Glassy phase}
\noindent
If the system is further cooled below the glass transition temperature $T_{g}$ a full relaxation of two time correlation functions cannot be observed anymore within the experimentally available time window. The calculation of the shear modulus of a material in this glassy state was extensively discussed by Williams. \cite{WiliamsI, WilliamsII} We briefly recap the physical picture considered by the authors there: The phase space of the system is divided into $N_{D}$ subsystems. Every subsystem is in equilibrium but between the domains the system is out of equilibrium. The probability density of the domain $a$ is given by $f_{a}(\Gamma)=s_{a}(\Gamma)\frac{\exp(-\beta H(\Gamma))}{Z_{a}}$ with $Z_{a}=\int d\Gamma s_{a}(\Gamma) \exp(-\beta H(\Gamma))$, where $\Gamma$ is the phase space coordinate, $H$ the Hamiltonian of the system and $s_{a}$ a switching function which is equal to unity if $\Gamma$ lies in the domain $a$ and zero otherwise. The probability distribution of the entire system is given by a composition of the single-domain distribution weighted with a nonequilibrium weight, i.e.\ $f(\Gamma)=\sum_{a=1}^{N_{D}} w_{a}f_{a}(\Gamma)$ and $\sum_{a=1}^{N_{D}} w_{a}=1$. Equation ($\ref{BG4}$) holds for the (equilibrium) subdomains only. The authors of this study further derived that the infinite frequency shear modulus of the system is given by
\begin{equation}
\label{BG6}
     G_{0}=G_{\infty,f} - \frac{V}{k_{B} T} \sum_{a=1}^{N_{D}} w_{a} \left( \left\langle \sigma_{xy}^2 \right\rangle_{a}  - \left\langle \sigma_{xy} \right\rangle_{a}^{2}  \right) \ ,
\end{equation}
where the subscript $f$ denotes the rule to average the Green-Born expression over the distribution $f$ and the subscript $a$ means an equilibrium average over the domain $a$.
This makes an accurate and meaningful calculation of the low frequency modulus for a glass sample a very subtle task as it would require to consider the single phase space domains. Under the assumptions that every simulation is sampling its own, single domain and that the set of prepared samples representatively reflects the distribution of the weights $w_{a}$, one could estimate ($\ref{BG6}$) by simple time averages. This approach was seemingly taken in [\onlinecite{Harro}] and in [\onlinecite{ilyin}]. However, as already pointed out in [\onlinecite{WiliamsI}] this method is very sensitive to the used time over which averages are taken.
Therefore, we mention an alternative approach, also presented in [\onlinecite{WiliamsI}]: the low frequency modulus is given by the Green-Born expression minus the stress-stress autocorrelation function $C(t)$ at $t=0$ (equation ($\ref{BG4}$)), which drops to a non-zero plateau value for a broad class of glass forming liquids. This means that the stress fluctuations relative to the frozen-in stresses in this domain are considered.
The autocorrelation is now calculated up to a cutoff time $t_{c}$ which is much larger then the relaxation time of the fast processes in the sample. Finally, the fluctuation term in ($\ref{BG4}$)) is corrected by $C(t_{c})$:
\begin{equation}
\label{BG7}
     G_{0}=G_{\infty} - \frac{V}{k_{B} T} \left( \left\langle \sigma_{xy}^2 \right\rangle  - \left\langle \sigma_{xy} \right\rangle^{2} - C(t_c) \right) \ .
\end{equation}
It should be noted that this procedure has the advantage that it is not sensitive to the chosen cutoff since $C(t)$ is almost constant for a broad time interval (meaning that ageing of the glass is negligible on the time scale of interest for the investigated model system). The physical picture behind this correction is the following: each individual glass sample contains frozen-in stresses, which are induced due to the initial conditions and preparation procedure of the sample. For instance, a fast cooling protocol pushes a liquid out of equilibrium very rapidly. This does not leave enough time for stress relaxation processes to occur leading to significant stresses in the glass sample which might not be present if a slower cooling rate would have been used. While these residual stresses are sometimes deliberately introduced during the manufacturing process \cite{residual} and influence mechanical/rheological measurements on individual glassy materials, they are not a characteristic property of the material but a remainder of its production process. Correcting for the frozen-in stresses by subtracting the plateau value of the stress-stress correlation function removes this contribution from the low frequency modulus such that $G_{0}$ remains a quantity which is characteristic for the material irrespectively of its history. As we will see in the next section, the IS stress may also bias the computation of the elastic moduli. Since, we want to investigate this effect in more detail, we focus on systems where corrections for the frozen-in stresses in the form of ($\ref{BG4}$) play a minor role only. This is the case for the considered system. While the self-intermediate scattering function does not decay to zero anymore, when the liquid passes the glass transition temperature \cite{model2d}, the stress-stress autocorrelation function $C(t)$ decays to values which are negligible for a potential correction according to equation ($\ref{BG4}$) (see fig.\ 2). The physical reason for this behavior might be, that stress relaxation events happen only in local rare events, but the global stress-stress correlation decays since very few rearrangements on the boundaries of the simulation box lead to a large change in the IS stress contribution as discussed in section I.

 \begin{figure}
 \includegraphics[scale=0.55]{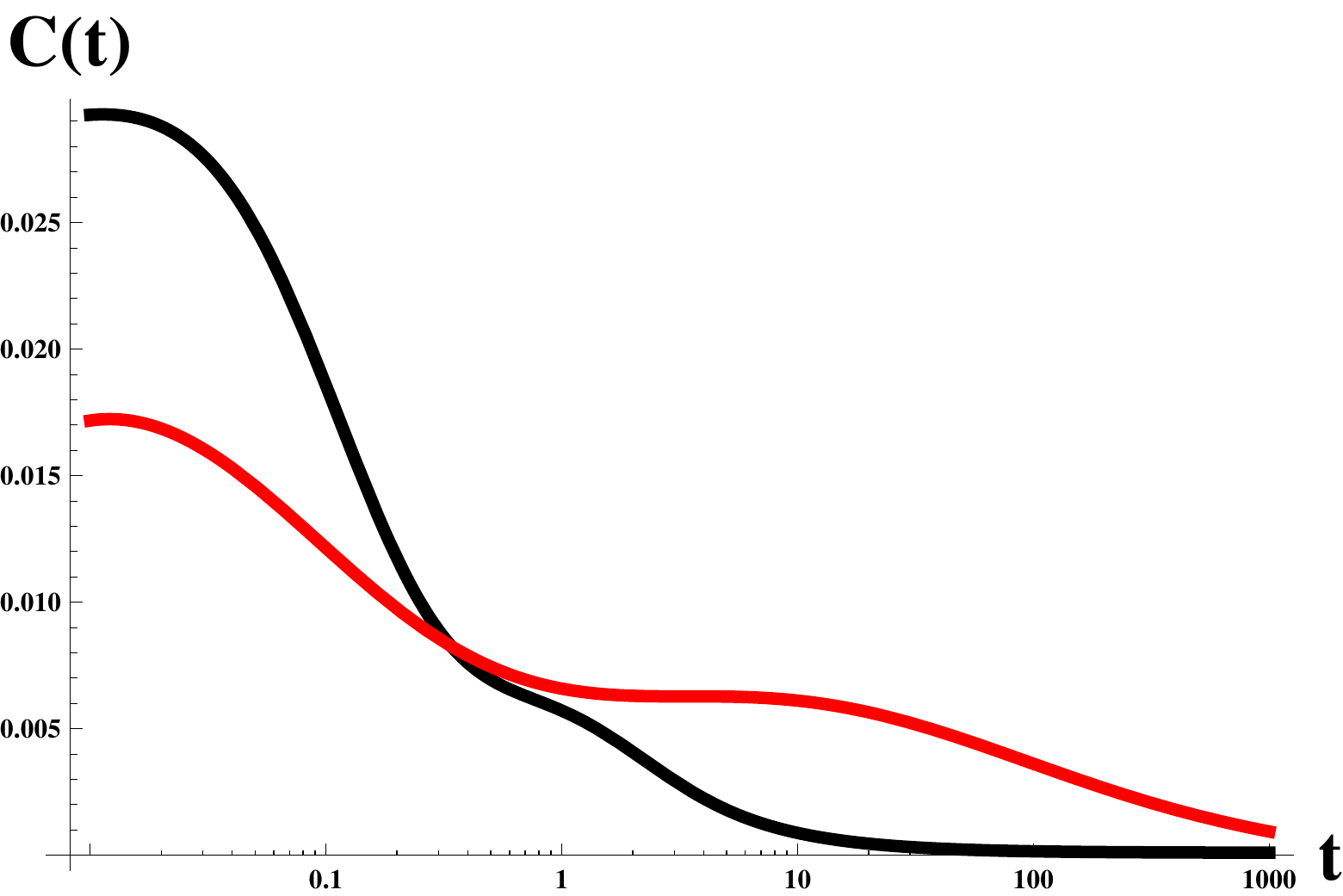}%
 \caption{\label{fig2} Stress-stress autocorrelation function $C(t)=\left \langle \sigma_{xy}(t)\sigma_{xy}(0)\right\rangle$ for high ($T=0.6$, black) and low temperature ($T=0.3$, red). While at $T=0.6$ the systems starts to develop a two step relaxation process, a pronounced plateau can be observed at $T=0.3$. Note, that for the system under consideration $C(t)$ drops to values close to zero even at temperatures below the nominated glass transition temperature $T \approx 0.35$. See text for an explanation. Numerical data have been fit by a stretched exponential $a \exp(-bt^{c})+d$ for long time scales.}
 \end{figure}

\subsection{Low temperature limit}
\noindent
The low temperature limit of the amorphous solid is of wide interest in the research community and subject to extensive scientific effort (for an overview see the corresponding sections in [\onlinecite{dynamic})]. In the following, we will extend our discussion on the calculations of the elastic properties to the low temperature regime. We will use the present considerations about the shear moduli as a tool to identify different mechanism contributing to the solidification process of a glassy material and clarifying the role of the IS stress played in it.

\begin{figure}[h!]
 \includegraphics[scale=0.65]{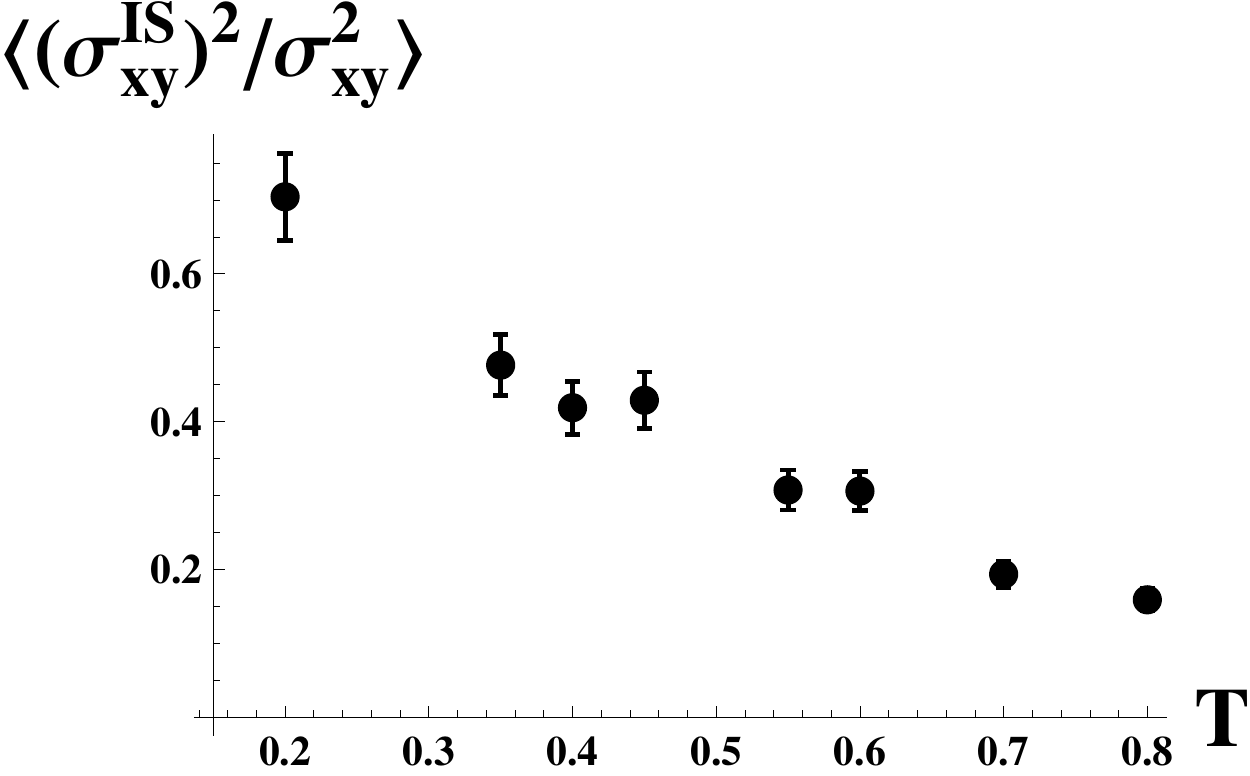}%
 \caption{\label{fig3} Fraction of stress variance stemming from the IS contribution, increasing over $50\%$ when the system enters the supercooled regime. At this point stress redistributions through thermal activation become so slow that the external constraint, set by the boundary conditions becomes non-negligible and requires a correction to the shear moduli.}
\end{figure}

\begin{figure}
 \includegraphics[scale=0.65]{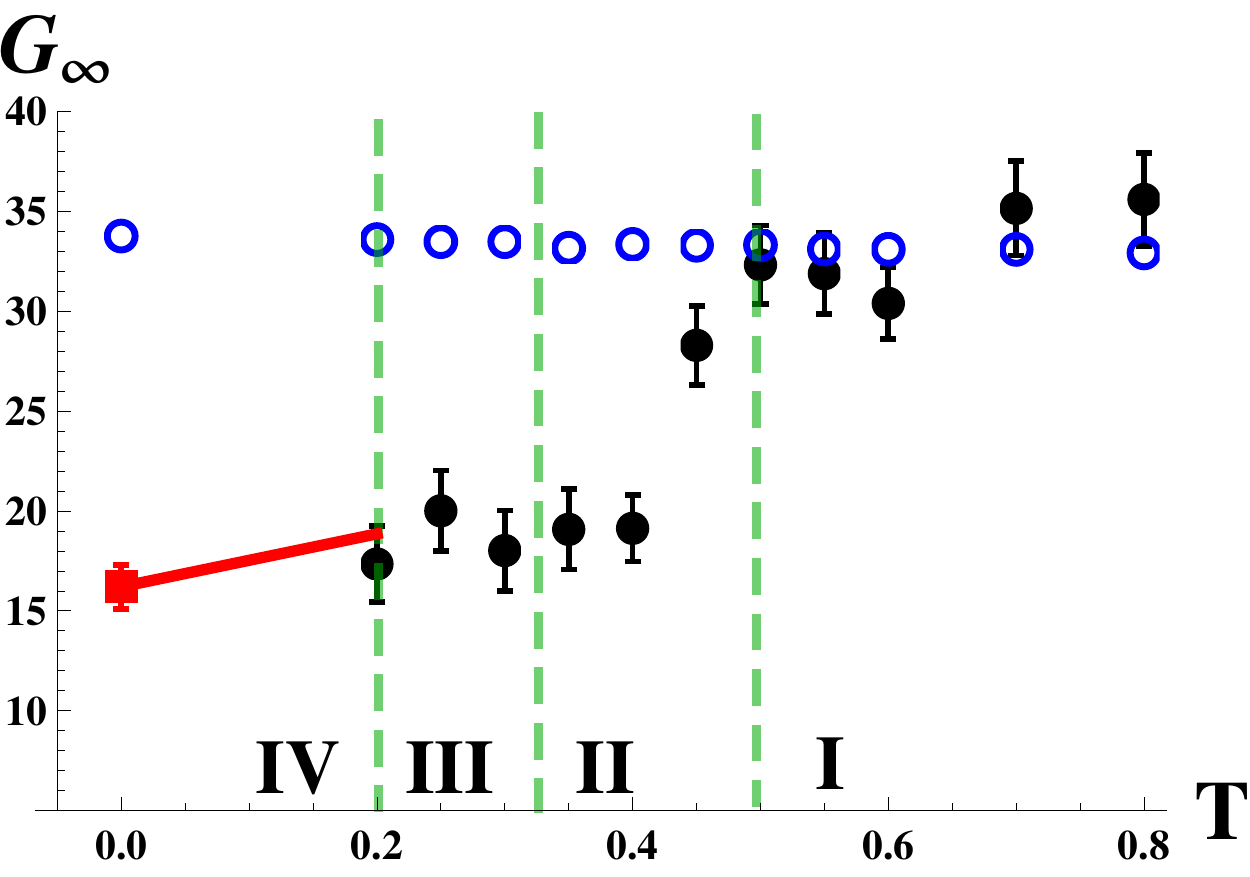}%
 \caption{\label{fig4} High frequency shear modulus (blue, open circles) according to equation ($\ref{BG3}$). Filled, black circles show simulation results for the fluctuation term from ensemble averaging with the proper IS corrections applied (see text). The red square is the zero temperature limit according to the first term in equation ($\ref{BG13}$). The red line corresponds to perturbation theory results for the fluctuation term of the Born-Green theory. The low frequency shear modulus is given by the difference between black and blue data points.  We distinguish four temperature regimes: Regime $I$ is the high temperature regime at which equations ($\ref{BG1}$) and ($\ref{BG5}$) hold equally to calculate the high frequency modulus. In regime $II$ the system is supercooled. The fluctuation does not fully cancel $G_{\infty}$ anymore meaning that the material develops an elastic behavior. Regime $III$ is the glassy state at which both frozen-in and IS stresses bias the calculation of the fluctuation term and are corrected to extract inherent material properties irrespectively of external constraints or history dependence. Regime $IV$ is the low temperature glass phase at which the fluctuation term can be computed with equation ($\ref{BG12}$)}
\end{figure}

\begin{figure}
 \includegraphics[scale=0.65]{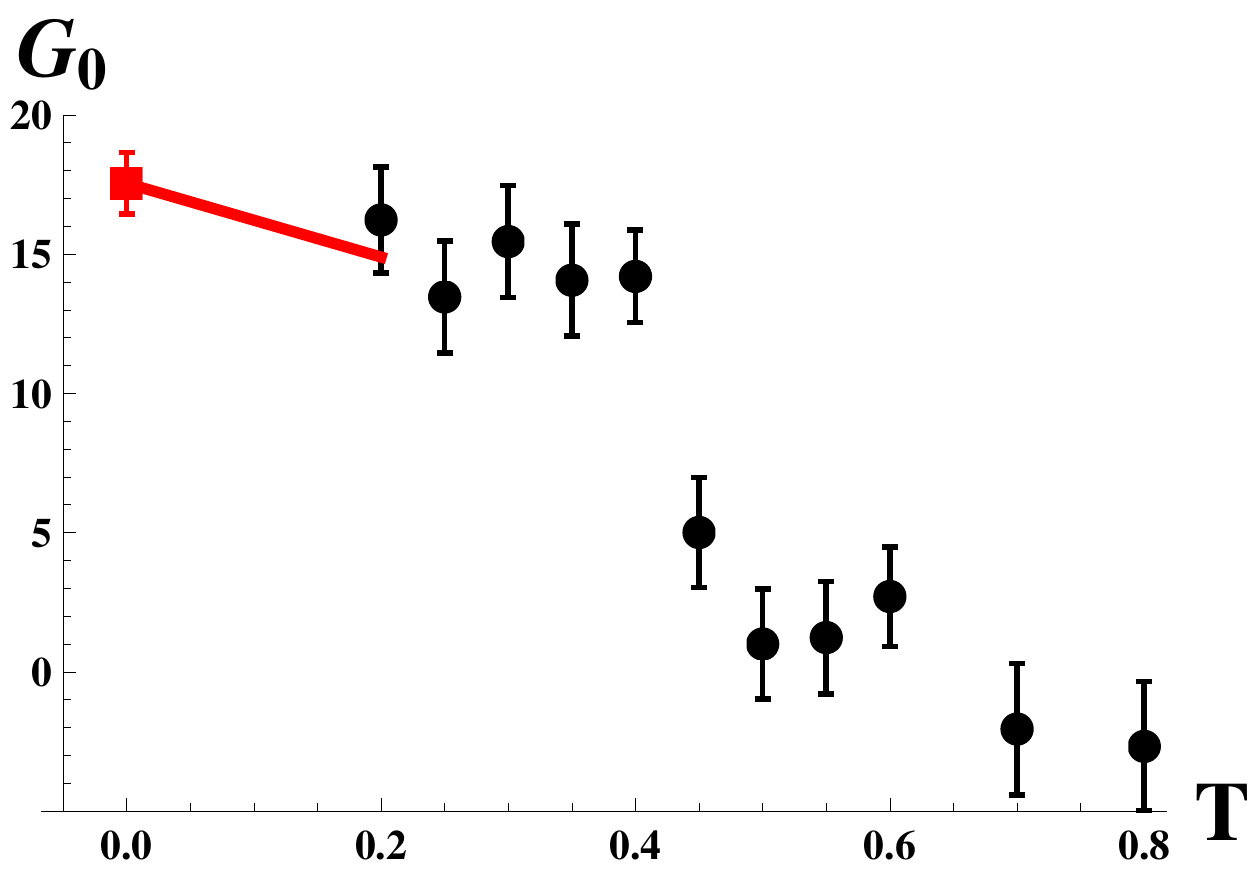}%
 \caption{\label{fig5} Zero frequency shear modulus obtained from subtracting the fluctuation term from the high frequency shear modulus according to equation ($\ref{BG4}$). Black circles show simulation results with proper IS corrections applied and the red square is the zero temperature limit according to the first term in equation ($\ref{BG12}$). The red line corresponds to perturbation theory results for the fluctuation term of the Born-Green theory. }
\end{figure}

For a subdomain $a$ which is assumed to be in equilibrium, we follow the calculation of Lutsko \cite{Lutsko} and compute the shear stress fluctuation in the canonical ensemble in the low temperature limit.
\begin{equation}
\label{BG8}
     \left\langle \sigma_{xy}^2 \right\rangle_{a}=\frac{1}{Z_{a}}  \int d\bm{r} s_{a} \sigma_{xy}^2 \exp \left( -\frac{\Phi}{k_{B}T}\right)     \ ,
\end{equation}
where $Z_{a}= \int d\bm{r} s_{a} \exp \left( -\frac{\Phi}{k_{B}T}\right)$ and $\Phi$ is the potential energy of the system. We expand both potential energy and stresses around its inherent structure, i.e.,
\begin{multline}
\label{BG9}
    \Phi=\left. \Phi \right|_{\bm{r}=\bm{r_{i}}^{IS}}+\left. \frac{\partial\Phi}{\partial r_{i,\alpha}}\right|_{\bm{r}=\bm{r_{i}}^{IS}}(r_{i,\alpha}-r_{i,\alpha}^{IS})+ \\ \frac{1}{2} \left.\frac{\partial^{2}\Phi}{\partial r_{i,\alpha}\partial r_{j,\beta}}\right|_{\bm{r}=\bm{r_{i}}^{IS}}(r_{i,\alpha}-r_{i,\alpha}^{IS})(r_{i,\beta}-r_{i,\beta}^{IS})+...     \ ,
\end{multline}
and
\begin{multline}
\label{BG10}
    \sigma_{xy}=\sigma_{xy}^{IS}+\left. \frac{\partial\sigma_{xy}}{\partial r_{i,\alpha}}\right|_{\bm{r}=\bm{r_{i}}^{IS}}(r_{i,\alpha}-r_{i,\alpha}^{IS})+ \\ \frac{1}{2} \left.\frac{\partial^{2}\sigma_{xy}}{\partial r_{i,\alpha}\partial r_{j,\beta}}\right|_{\bm{r}=\bm{r_{i}}^{IS}}(r_{i,\alpha}-r_{i,\alpha}^{IS})(r_{i,\beta}-r_{i,\beta}^{IS})+...     \ ,
\end{multline}
where we have used the summation convention for indices. In the following, we will use $\Phi_{n}^{IS}$ and $\sigma_{xy,n}^{IS}$ as a short notation, where the superscript means that the expression has to be evaluated at the inherent structure configuration and the number in the subscript refers to the $n$-th derivative. Transforming to re-scaled coordinates $r_{i,\alpha}-r_{i,\alpha}^{IS}=(k_{B}T)^{1/2}r'_{i,\alpha}$ and inserting ($\ref{BG9}$) and ($\ref{BG10}$) into equation ($\ref{BG8}$) yields
\begin{multline}
\label{BG11}
   \int s_{a} \left( \sigma_{xy}^{IS}+ \sqrt{k_{B}T}\sigma_{xy,1}^{IS} r'+ ... \right)^{2} \\
      \exp \left( -\frac{1}{2}\Phi_{2}^{IS}r'r' \right)  \exp \left(-\frac{\sqrt{k_{B}T}}{6}\Phi_{3}^{IS}r'r'r' -...  \right) d\bm{r'} / \\ \int s_{a} \exp \left( -\frac{1}{2}\Phi_{2}^{IS}r'r' \right)  \exp \left(-\frac{\sqrt{k_{B}T}}{6}\Phi_{3}^{IS}r'r'r' -...  \right) d\bm{r'} \ .
\end{multline}
Expanding the second exponential factor in powers of $k_{B}T$ (in both denominator and enumerator), all integrals are Gaussian integrals, which are solvable analytically. Note that odd moments of these Gaussian integrals vanish. Finally, expanding the quotient in equation ($\ref{BG11}$) in powers of $k_{B}T$ leads directly to the following result for the stress-stress fluctuation in the low temperature limit:
\begin{multline}
\label{BG12}
    \frac{V}{k_{B}T} \left\langle \sigma_{xy}^2 \right\rangle_{f} \approx \\ V \sum_{a} w_{a} \left(\frac{(\sigma_{xy}^{IS,a})^2}{k_{B}T} + A + B k_{B}T + \mathcal O((k_{B}T)^{2}) \right)     \ ,
\end{multline}
where $\sigma_{xy}^{IS,a}$ is the inherent structure contribution from particles in subdomain $a$ and the subscript $f$ has the same meaning as in equation (($\ref{BG6}$)). At this point we have to make further assumptions in order to estimate the low temperature limit of the fluctuation term. Again we are left with the problem of identifying the different subdomains in order to estimate the weights $w_{a}$ and to sum over $\left\langle \sigma_{xy}^2 \right\rangle_{a}$ accordingly. We will make the previously mentioned assumption that the prepared samples reflect the distribution of these weights and that each simulation predominantly samples its own single domain such that the total inherent structure stress of one simulation, $\sigma_{xy}^{IS}$ approximates $\sigma_{xy}^{IS,a}$.
Therefore, the low temperature limit of the fluctuation term is estimated
by ensemble averaging over statistically independant starting configurations.
Hence, the quantity $A$ is the linear term in the perturbation expansion and given by
\begin{multline}
\label{BG13}
    A = \left( \frac{\partial \sigma^{IS}_{xy}}{\partial r_{i\alpha}} \frac{\partial \sigma_{xy}^{IS}}{\partial r_{j\beta}}\right) \langle\langle r'_{i\alpha}r'_{j\beta} \rangle\rangle \\ -\frac{1}{3}\sigma^{IS}_{xy}\left( \frac{\partial^{3}\Phi}{\partial r_{i\alpha}\partial r_{j\beta}\partial r_{k\gamma}} \frac{\partial \sigma_{xy}^{IS}}{\partial r_{l\delta}}\right) \langle\langle r'_{i\alpha}r'_{j\beta}r'_{k\gamma}r'_{l\delta} \rangle\rangle \\ + \sigma^{IS}_{xy}\left( \frac{\partial^{2}\sigma^{IS}_{xy}}{\partial r_{i\alpha}\partial r_{j\beta}}  \right) \langle\langle r'_{i\alpha}r'_{j\beta} \rangle\rangle \ .
\end{multline}
The contributions to the linear term $B$ are discussed later.
We have introduced the Gauss bracket of a function $g$ which is defined as follows $\langle \langle g(\bm{r}') \rangle \rangle= \frac{1}{c} \int d\bm{r}' g(\bm{r}') \exp \left( -\Phi_{2}^{IS}\bm{r}'\bm{r}' \right)$, where $c$ is a normalization constant such that $\langle \langle 1 \rangle \rangle=1$ holds. Note, that
\begin{equation}
\label{BG14}
   \langle \langle r_{i\alpha}r_{j\beta} \rangle \rangle= \left( (\Phi_{2}^{IS})_{i\alpha,j\beta} \right)^{-1} \,
\end{equation}
where the right hand side is nothing but the inverse of the Hessian matrix. The Hessian matrix has zero eigenvalues due to the fact the the net force on the system is zero and is therfore not invertible. A commonly used solution to this issue is, physically holding one particle fixed (i.e.\ excluding it from the sums in all calculations), which has no effect on the free energy. \cite{hoover} All higher moments like $\langle \langle r_{i\alpha}r_{j\beta}r_{k\gamma}r_{l\delta} \rangle \rangle$ can be traced back to ($\ref{BG14}$) using Wick's theorem, e.g.:
 \begin{multline}
\label{BG114b}
     \langle\langle r_{i\alpha}r_{j\beta}r_{k\gamma}r_{l\delta} \rangle\rangle=  \langle \langle r_{i\alpha}r_{j\beta} \rangle \rangle \langle \langle r_{k\gamma}r_{l\delta} \rangle \rangle  + \\ \langle \langle r_{i\alpha}r_{k\gamma} \rangle \rangle \langle \langle r_{j\beta}r_{l\delta} \rangle \rangle + \langle \langle r_{i\alpha}r_{l\delta} \rangle \rangle \langle \langle r_{j\beta}r_{k\gamma} \rangle \rangle \ .
\end{multline}
 We find that the first term in equation ($\ref{BG12}$) is exactly the inherent structure contribution to the stress fluctuations. Since these fluctuations are essentially temperature independent, this term would lead to a divergence of the low frequency shear modulus. This means that, under the constraint of the given containment (or the boundary conditions of the simulation box) the particles possibly cannot be packed in a way which zeros the total stress as discussed in the previous section. We note that this constraint is imposed on the entire box so it does not affect the results of local elastic quantities where the subvolume is embedded in a larger system (e.g.\ see [\onlinecite{JLB}] or [\onlinecite{pab}]). At high temperatures this term also does not play an important role: for formal reasons due to the {1/T} prefactor, for physical reasons due to permanent stress redistributions activated by thermal motion occurring at high temperatures. These stress fluctuations are usually much larger than its IS contribution (see figure ($\ref{fig3}$)).
We identify the IS contribution to the shear stress fluctuation as \textit{geometric} frustration of the system, accompanying the \textit{kinematic} frustration, which has been discussed in the previous section on the nonequilibrium glassy state. Again, this contribution is not an inherent material property but reflects external constraints on the relaxation dynamics becoming significant at low temperatures. The second term in ($\ref{BG12}$) is temperature independent and the benefit of equation ($\ref{BG13}$) is that, we can read off how the stress fluctuations obtained from a simulation should be corrected to extract the true zero temperature properties of the material when frozen-in stresses can be neglected. Alternatively, the material properties can be calculated from the first term in ($\ref{BG13}$) which does not contain the IS shear stress itself but only its derivative. The latter is nonzero irrespectively of any constraint on the system. This term has already been deduced in [\onlinecite{Lutsko}], however it has so far not been numerically tested for amorphous systems to the best of our knowledge. Even though equation ($\ref{BG12}$) is strictly speaking only valid in equilibrium at low temperatures, we propose to correct the zero frequency shear modulus even in the glassy regime, since the inherent structure stress plays a predominant role already at this temperature range as can be seen in figure $\ref{fig3}$.
The second term, $B$, in equation ($\ref{BG12}$) describes the slope at which the fluctuation term departs from its zero temperature limit. Among all terms contributing to this order of $(k_{B}T)$ we neglect those which contain IS contributions (see Appendix B) and obtain the temperature dependence of the fluctuation term close to $T=0$. \\
As a conclusion, we have obtained a detailed picture of the computation of the high and low frequency shear moduli of a glass forming system:
While in the high temperature regime the fluctuation term cancels the infinite frequency shear modulus, it decreases in the supercooled regime. This means the the glass forming material develops an elastic behavior. Due to the non-thermal IS contribution the fluctuation term would increase again upon further cooling in the glassy phase. As we correct for this contribution, we observe a small decrease of the fluctuation term towards the low temperature limit (see figure $\ref{fig4}$). The difference between $G_{\infty}$ and the fluctuation term is depicted in figure 5. At high temperatures $G_{0}$ is essentially zero meaning that the fluid does not support stresses on a long time scale. Entering the supercooled regime the material behaves elastically, which is mirrored by an increase of the zero frequency modulus. As the temperatures is reduced further this elastic behavior becomes more pronounced (see figure $\ref{fig5}$).

%
%
\label{IK10}
%
%

\section{Conclusion}
\noindent
In this note we discussed the origin of the inherent structure stress and traced it back to the particular choice of boundary conditions. In view of equation ($\ref{IK4}$) the IS stress can be understood as the net forces penetrating the boundary of the simulation box. Notably, the formula ($\ref{IK4}$) exactly coincides with the so-called ``Method of Planes'' definition of the stress evaluated at the boundaries of the simulation box. \cite{mop1,mop2} This analytic expression allowed us to explain the scaling of the IS stress with the system size (see appendix). Moreover, it might serve as starting point for understanding other properties of the IS stress, e.g.\ its temperature independence or its absolute magnitude. 
Finally, we would like to address the question which physical meaning lies behind the IS stress. It is obvious that in a sufficiently low temperature regime the configurational part of the stress tensor becomes dominant over the kinetic contribution. Starting from the Green-Kubo formula for the viscosity\cite{zwanzig}, $\eta=\frac{V}{k_{B}T}\int_{0}^{\infty}\langle \sigma_{xy}(t)\sigma_{xy}(0) \rangle dt$ , it has been noted that the IS stress autocorrelation describes the onset of highly viscous behavior as a liquid enters its supercooled regime.\cite{Harro} Hence, as the system explores the potential energy landscape, hopping from one minimum to another, the underlying IS stress fluctuations determine its viscosity in the linear response regime. In view of equation ($\ref{IK4}$) this means that a rearrangement of forces at the boundary of the system contains enough information to characterize the inherent structure which the system currently resides in. However, two caveats seem to come along with this notion. First, the IS stress itself does not provide any obvious information on which time scale its autocorrelation will decay or at which frequency the hopping between the energy minima occurs. It is merely associated with a geometric frustration which is imposed on the system by choosing particular boundary conditions. Secondly, one should not conclude from ($\ref{IK4}$) that cooperative rearrangements taking place at the transition from one minimum to another primarily happen at the boundary of the system. They occur anywhere in the system but the energy minimization zeroes all net forces on particles locally, leading to a redistribution of the boundary forces. In this sense the boundary conditions introduce a frustration which is present everywhere in the system.
In view of the previous discussion the following, overall picture for the role of the IS stress emerges:
the IS stresses and the redistributions of forces might amount to a macroscopic, scalar observable which characterizes the inherent structure such that its autocorrelation mirrors the path of the system through its configuration space, but does not contribute to the viscosity via any associated timescale as it does not contain dynamical information itself.
However, it affects the elastic properties of the system as it leads to a diverging contribution to the fluctuation term of the Born-Green theory. The reason for this behavior is that the IS contribution to the fluctuation term is of a non-thermal origin. If this term is divided by $k_{B}T$, it leads to a divergence of the fluctuation term and therefore to results for the elastic moduli which are biased by the external constraint imposed on the system. Since every rheological (or mechanical) measurement goes together with a change of the boundary of the system and as there is no experimental indication for a low-temperature divergence of the infinite frequency shear moduli in glassy systems \cite{Zhang,jiang}, it seems appropriate to remove the IS contribution to the shear modulus in order to correct for external constraints in the preparation procedure of the IS.


%
%

%

\begin{acknowledgments}
  We thank Hans-Christian {\"O}ttinger and Jean-Louis Barrat for insightful discussions and gratefully acknowledge the Swiss National Science Foundation for providing funding under Grant No.\ 200021\_134626.
\end{acknowledgments}


\appendix

\section{Model systems and simulation details}

The chosen model system and simulation details essentially correspond to those used by Abraham and Harrowell \cite{Harro} investigating glass forming model systems in 2D \cite{model2d} and 3D \cite{model3d}. Molecular dynamics simulations were carried out at fixed volume, temperature and particle number, using the LAMMPS package. \cite{LAMMPS} We used equimolar, binary mixtures of particles interacting via a purely repulsive potential in 2D, $\phi_{ij}(r)=\epsilon \left( \frac{ \sigma_{ij} }{r} \right)^{12}$, with $\epsilon=1$, $\sigma_{11}=1$, $\sigma_{12}=1.2$,$\sigma_{22}=1.4$ and a Lennard-Jones potential in 3D, $\phi_{ij}(r)=4 \epsilon \left( (\frac{ \sigma_{ij} }{r})^{12} - (\frac{ \sigma_{ij} }{r})^{-6} \right)$ with $\epsilon=1$, $\sigma_{11}=1$, $\sigma_{12}=1.1$,$\sigma_{22}=1.2$. A cutoff $r_{c}$ was used at $4.5 \sigma_{11}$ and the potential were shifted such that they vanished at $r_{c}$. The mass of the particles were $m_{1}=1$ and $m_{2}=2$ and the used densities were $\rho=0.747$ in 2D and $\rho=0.75$ in 3D. Both systems were simulated at $T=0.5$ which is below the freezing temperature of both systems and the particle numbers were chosen to be $N=512,1024,2048,4096,8192$ in 2D and $N=1024,2048,4096,8192$ in 3D. Reduced units are used for length $L^{*}=L/\sigma_{11}$, temperature $T^{*}=k_{B}T/\epsilon$, energy, $E^{*}=E/\epsilon$ and pressure $P^{*}=P\sigma^{3}/\epsilon$. Throughout all simulations, periodic boundary conditions are applied.

\section{Perturbation calculations}

In the following we list the terms which contribute to the linear order term in equation ($\ref{BG12}$), $B=\sum_{i=1}^{7}B_{i}$.

\begin{align}
    B_{1}&=\frac{1}{24} \sigma_{xy,\alpha_{1}} \sigma_{xy,\alpha_{2}} \Phi_{\alpha_{3}\alpha_{4}\alpha_{5}\alpha_{6}}   \langle \langle r'_{\alpha_{1}} r'_{\alpha_{2}} \rangle \rangle \left< \left< \prod_{i=3}^{6} r'_{\alpha_{i}} \right> \right>  \notag \\
    B_{2}&=\frac{-\sigma_{xy,\alpha_{1}} \sigma_{xy,\alpha_{2}}}{72}  \Phi_{\alpha_{3}\alpha_{4}\alpha_{5}}\Phi_{\alpha_{6}\alpha_{7}\alpha_{8}}  \notag \\ & \left< \left< \prod_{i=3}^{8} r'_{\alpha_{i}} \right> \right> \langle \langle r'_{\alpha_{1}} r'_{\alpha_{2}} \rangle \rangle \notag \\
    B_{3}&=\frac{\sigma_{xy,\alpha_{1}} \sigma_{xy,\alpha_{2}}}{72}  \Phi_{\alpha_{3}\alpha_{4}\alpha_{5}}\Phi_{\alpha_{6}\alpha_{7}\alpha_{8}}   \left< \left< \prod_{i=1}^{8} r'_{\alpha_{i}} \right> \right>  \notag \\
    B_{4}&=- \frac{1}{24}\sigma_{xy,\alpha_{1}} \sigma_{xy,\alpha_{2}} \Phi_{\alpha_{3}\alpha_{4}\alpha_{5}\alpha_{6}}   \left< \left< \prod_{i=1}^{6} r'_{\alpha_{i}} \right> \right>  \notag \\
    B_{5}&=\frac{1}{4}\sigma_{xy,\alpha_{1}\alpha_{2}} \sigma_{xy,\alpha_{3}\alpha_{4}} \left< \left< \prod_{i=1}^{4} r'_{\alpha_{i}} \right> \right>  \notag \\
    B_{6}&=-\frac{1}{6} \sigma_{xy,\alpha_{1}} \sigma_{xy,\alpha_{2}\alpha_{3}} \Phi_{\alpha_{4}\alpha_{5}\alpha_{6}}  \left< \left< \prod_{i=1}^{6} r'_{\alpha_{i}} \right> \right>  \notag \\
    B_{7}&=\frac{1}{3} \sigma_{xy,\alpha_{1}} \sigma_{xy,\alpha_{2}\alpha_{3}\alpha_{4}} \left< \left< \prod_{i=1}^{4} r'_{\alpha_{i}} \right> \right> \notag
\end{align}

Note that the calculation of this first order term is a somewhat lengthy and complicated task: Firstly, higher derivatives of the total potential $\Phi$ and of the Irving-Kirkwood expression of the stress are required. However, these can be computed in a straightforward way for simple pair potentials (like the purely repulsive soft sphere potential used in this study) by using software which is capable of performing symbolic computations.
Secondly, some of these objects (e.g.\ the fourth derivative of the potential $\Phi_{\alpha_{1}\alpha_{2}\alpha_{3}\alpha_{4}}$) contains a very large number of entries. In the present study we investigate systems with $512$ particles in two dimensions meaning that this tensor has approximately $(10^{3})^{4}=10^{12}$ entries. Summing over this number of entries and even storing the object itself seems to be an intractable task. However, the majority of entries are zero, since the derivative with respect to three (or more) different particle coordinates vanishes for a pair potential. Additionally, the sequence of the performed derivatives does not influence its value according to Schwartz's theorem.
A third problem arises due the use of Wick's theorem: the term $B_{3}$ requires the calculation of the eighth moment. Splitting this in all possible pair combinations yields in total $105$ terms but the problem can be simplified using symmetry properties of the involved objects. In the following we provide further details on how one of the terms is calculated in practice. The other terms can be handled in a completely analogous way.
For instance the term $B_{4}$ can be split as follows:
\begin{multline}
\label{appb1}
    B_{4}=- \frac{1}{24}\sigma_{xy,\alpha_{1}} \sigma_{xy,\alpha_{2}} \Phi_{\alpha_{3}\alpha_{4}\alpha_{5}\alpha_{6}} \\ \left(  \left< \left< r'_{\alpha_{1}} r'_{\alpha_{2}} \right> \right> \left< \left< r'_{\alpha_{3}} r'_{\alpha_{4}} r'_{\alpha_{5}} r'_{\alpha_{6}} \right> \right> \right. + \\  \left. \left< \left< r'_{\alpha_{1}} r'_{\alpha_{3}} \right> \right> \left< \left< r'_{\alpha_{2}} r'_{\alpha_{4}} r'_{\alpha_{5}} r'_{\alpha_{6}} \right> \right> + ...  \right) \ .
\end{multline}
The first term of the sum cancels $B_{1}$. Contracting over $\alpha_{1}$ and introducing the abbreviation $T_{\alpha_{3}}=\sigma_{xy,\alpha_{1}} \left< \left< r'_{\alpha_{1}} r'_{\alpha_{3}}\right> \right>$, this can be rewritten
\begin{multline}
\label{appb2}
    B_{4}=- \frac{1}{24} \sigma_{xy,\alpha_{2}} \Phi_{\alpha_{3}\alpha_{4}\alpha_{5}\alpha_{6}} \\ \left(  T_{\alpha_{3}} \left< \left< r'_{\alpha_{2}} r'_{\alpha_{4}} r'_{\alpha_{5}} r'_{\alpha_{6}} \right> \right> \right. + \\  \left. T_{\alpha_{4}} \left< \left< r'_{\alpha_{2}} r'_{\alpha_{3}} r'_{\alpha_{5}} r'_{\alpha_{6}} \right> \right> + ...  \right) \ .
\end{multline}
Since the sequence of the derivatives of $\Phi$ does not matter all terms in the brackets are the same which leads to
\begin{multline}
\label{appb3}
    B_{4}=- \frac{4}{24} \sigma_{xy,\alpha_{2}} \Phi_{\alpha_{3}\alpha_{4}\alpha_{5}\alpha_{6}}  \left(  T_{\alpha_{3}} \left< \left< r'_{\alpha_{2}} r'_{\alpha_{4}} r'_{\alpha_{5}} r'_{\alpha_{6}} \right> \right> \right) \ .
\end{multline}
Employing Wick's theorem another time, i.e.:
\begin{multline}
\label{appb4}
 \left< \left< r'_{\alpha_{2}} r'_{\alpha_{4}} r'_{\alpha_{5}} r'_{\alpha_{6}} \right> \right>= \left< \left< r'_{\alpha_{4}} r'_{\alpha_{6}} \right> \right>\left< \left< r'_{\alpha_{2}}  r'_{\alpha_{5}} \right> \right>+ \\ \left< \left< r'_{\alpha_{4}} r'_{\alpha_{5}}  \right> \right>\left< \left< r'_{\alpha_{2}} r'_{\alpha_{6}} \right> \right>+\left< \left< r'_{\alpha_{5}} r'_{\alpha_{6}} \right> \right>\left< \left< r'_{\alpha_{2}} r'_{\alpha_{4}} \right> \right> \,
\end{multline}
 and using the permutability of indices again, we conclude:
\begin{multline}
\label{appb5}
B_{4}=- \frac{12}{24} T_{\alpha_{3}} T_{\alpha_{4}} \Phi_{\alpha_{3}\alpha_{4}\alpha_{5}\alpha_{6}}   \left< \left< r'_{\alpha_{5}} r'_{\alpha_{6}} \right> \right>  \ .
\end{multline}
This expression can be handled numerically bearing in mind that only those entries of $\Phi_{\alpha_{3}\alpha_{4}\alpha_{5}\alpha_{6}}$ are non-zero for which at least three particle indices are identical. With these simplifications the first order contributions $B_{1}$ to $B_{7}$ can be calculated. We note that, even though the slope of temperature dependence is in good agreement with the medium temperature data the sample to sample fluctuations of the first order term $B$ is much larger than those of the zeroth order term $A$.

\section{Volume dependence}

In this appendix we provide an intuitive explanation for the scaling behavior of the IS stress variance.
We start by briefly recapitulating a simple density scaling argument for plastic flow of amorphous solids made in [\onlinecite{Itamar2}] and [\onlinecite{Itamar1}], which was used previously \cite{Harro} to explain the density scaling of the IS stress. We consider a purely repulsive pair potential $\phi(r) = (\sigma/r)^{n}$ and the force magnitude between particle $i$ and $j$ is given by $F_{ij}=\frac{\partial \phi(r_{ij})}{\partial r_{ij}} \sim r_{ij}^{-n-1}$. If we denote the probability distribution of distances $r$ at density $\rho$ by $p(r,\rho)$, the mean distance is given $r_{0}(\rho)=\int r p(r,\rho) dr$. Assuming that $p(r,\rho)$ is strongly peaked around a characteristic distance, it can be estimated that $r_{0} \sim \sigma / \rho^{1/d}$ in $d$ dimensions. Therefore, we assume the interaction between particle $i$ and $j$ to be determined solely by density of the system but to be independent of its size. Summing over the index $j$ in ($\ref{IK4}$), we obtain

\begin{equation}
\label{IK5}
     \sigma_{x y}^{IS} \approx  - \frac{L}{V} \sum_{i=1}^{N_{b}} F_{i,y}^{x=L/2}   \ ,
\end{equation}

where $F_{i,y}^{x=L/2}$ is the $y$-component of the net force acting on a particle $i$ for which $x_{i} \lesssim L/2$ exerted by the periodic images of the particles residing in the vicinity of the $x_{j}=-L/2$ boundary. $N_{b}$ is the number of particles in the boundary layer of the simulation box i.e.\ $(x_{i},y_{i}) \in \left[L/2-r_{c},L/2\right] \times \left[-L/2,L/2\right]$, where $r_{c}$ is the interaction range of the potential. We also multiplied by a factor of two to account for the reversed situation where particle $i$ resides at $x_{i} \gtrsim -L/2$. Clearly, the average over different configurations vanishes, $\langle \sigma_{x y}^{IS} \rangle \approx 0$ but we find for the average magnitude of the IS stress:

\begin{equation}
\label{IK6}
     \left< (\sigma_{x y}^{IS})^2 \right> \approx   \frac{L^2}{V^2} \left< \sum_{i=1}^{N_{b}} F_{i,y}^{x=L/2} \sum_{j=1}^{N_{b}} F_{j,y}^{x=L/2} \right> \ .
\end{equation}
The sum in ($\ref{IK6}$) extends over particles which are in close vicinity of each other. Therefore, the forces $F_{i,y}^{x=L/2}$ cannot be regarded to be uncorrelated but might be assumed to be weakly anti-correlated over distance of a few particle diameters. From a physical point of view, this is perfectly sensible, as will be made clear by the following intuitive argument. Assume, a particle $i_{1}$ is subjected to a strong, positive force ($F_{i_{1},y}^{x=L/2}= a > 0$). Then, a particle $i_{2}$, which is not to far away, should push with the same force $F_{i_{1},y}^{x=L/2} \approx -a$ in the reverse direction in order to maintain the mechanical stability in which the inherent structure configuration resides per definition. It is clearly not possible that all forces perfectly cancel in that manner. We denote the number of unbalanced forces by $N_{c}$ which should scale as the number of particles in the corner of the simulation box i.e.\ $N_{c} \sim r_{c}^{2}L^{d-2}$. Applying this argument to equation ($\ref{IK6}$) leads to

\begin{multline}
\label{IK7}
     \left< (\sigma_{x y}^{IS})^2 \right> \approx   \frac{L^2}{V^2} \left< \sum_{i=1}^{N_{b}} F_{i,y}^{x=L/2} \sum_{j=1}^{N_{b}} F_{j,y}^{x=L/2} \right> \\ \approx \frac{L^2}{V^2} \left< \sum_{i=1}^{N_{c}} \left( F_{i,y}^{x=L/2} \right)^2 \right> \ ,
\end{multline}

where the remaining forces in ($\ref{IK7}$) are uncorrelated. Invoking the central limit theorem to the presumably independent remaining forces results in the following scaling estimate
\begin{equation}
\label{IK8}
    \langle (\sigma_{x y}^{IS})^2 \rangle \approx \frac{L^2}{V^2} \left< \sum_{i=1}^{N_{c}} \left( F_{i,y}^{x=L/2} \right)^2 \right> \sim \frac{L^2}{V^2}L^{d-2} \sim L^{-d} \ ,
\end{equation}

which interestingly reveals the IS stress fluctuations to be a $V^{-1}$ effect although shown to arise from a contribution originating from a constraint imposed on the boundary of the box. This scaling behavior coincides with the one numerically found by the authors of [\onlinecite{Harro}]. \\

\bibliography{if_IS_ref}

\end{document}